\documentclass[conference]{IEEEtran}
\IEEEoverridecommandlockouts
\usepackage{cite}
\usepackage{amsmath,amssymb,amsfonts}
\usepackage{algorithmic}
\usepackage{graphicx}
\usepackage{textcomp}
\usepackage{xcolor}
\usepackage{subcaption}
\usepackage{url}
\usepackage{array}
\usepackage{multirow}
\usepackage{booktabs}
\def\BibTeX{{\rm B\kern-.05em{\sc i\kern-.025em b}\kern-.08em
    T\kern-.1667em\lower.7ex\hbox{E}\kern-.125emX}}
\begin{document}

\title{Proactively Detecting Threats: A Novel Approach Using LLMs
}

\author{\IEEEauthorblockN{Aniesh Chawla\IEEEauthorrefmark{1}, Udbhav Prasad\IEEEauthorrefmark{1}}
\IEEEauthorblockA{\IEEEauthorrefmark{1}
California, USA \\
\{chawla.aniesh, udbhav523\}@gmail.com}
\IEEEauthorblockA{\IEEEauthorrefmark{1}These authors contributed equally to this work.}}

\maketitle

\begin{abstract}
Enterprise security faces escalating threats from sophisticated malware, compounded by expanding digital operations. This paper presents the first systematic evaluation of large language models (LLMs) to proactively identify indicators of compromise (IOCs) from unstructured web-based threat intelligence sources, distinguishing it from reactive malware detection approaches. We developed an automated system that pulls IOCs from 15 web-based threat report sources to evaluate six LLM models (Gemini, Qwen, and Llama variants). Our evaluation of 479 webpages containing 2,658 IOCs (711 IPv4 addresses, 502 IPv6 addresses, 1,445 domains) reveals significant performance variations. Gemini 1.5 Pro achieved 0.958 precision and 0.788 specificity for malicious IOC identification, while demonstrating perfect recall (1.0) for actual threats. 
\end{abstract}


\section{Introduction}

\begin{figure}
    \centering
    \includegraphics[width=0.9\linewidth]{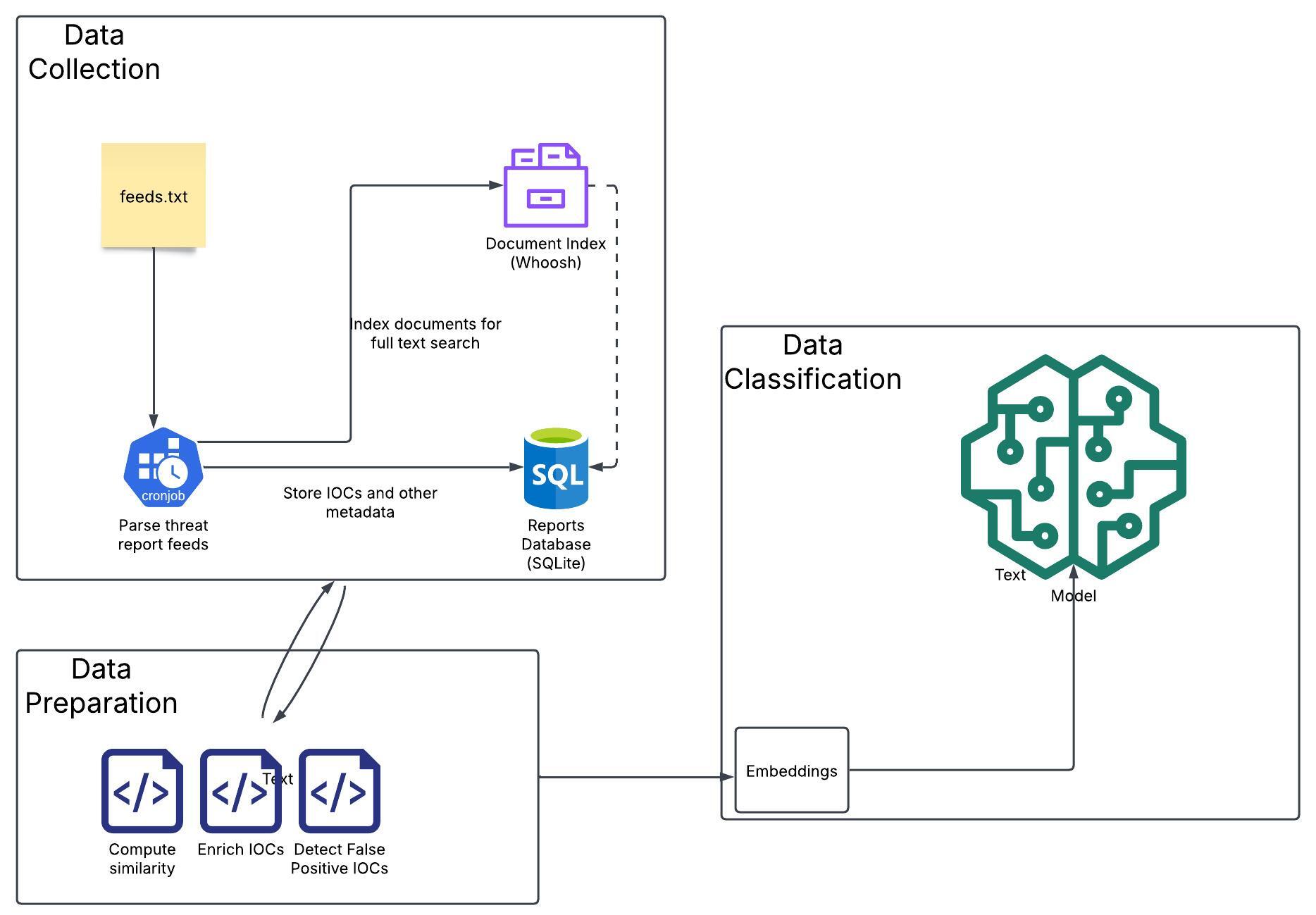}
    \caption{System to Evaluate threat indicators from the web}
    \label{fig:overall-system}
\end{figure}

The threat landscape for enterprise malware is growing increasingly sophisticated, exacerbated by expanded digital operations through remote work and cloud services, which broaden attack surfaces. The FBI’s 2023 Internet Crime Report \cite{ic3_annual_report_2023} highlights significant financial impacts, including \$2.9 billion lost to Business Email Compromise scams, a 38\% rise in investment fraud to \$4.57 billion, and \$59.6 million in ransomware costs. Similarly, IBM’s 2024 Cost of a Data Breach Report \cite{ibm_cost_data_breach_2024} reveals a 10\% increase in average breach costs, now \$4.88 million, with organizations that use AI for incident prevention, saving an average of \$2.2 million. These factors emphasize the urgent need for advanced AI-driven security tools. This paper explores the potential of large language models (LLMs) for proactive identification of Indicators of Compromise (IOCs) that enterprises can integrate into detection systems preemptively. We developed an automated system (Fig.\ref{fig:overall-system}) that collects threat data from reliable sources, parses the information into a database, and evaluates LLM models to classify whether the data contains IOCs, aiming to shift from reactive to proactive cybersecurity measures.

\section{Related Work}

\subsection{Benchmark Datasets and Evaluation Frameworks}

Recent efforts have focused on creating standardized datasets for evaluating cybersecurity AI systems. CyberMetric \cite{cybermetric} established a benchmark Q\&A dataset for evaluating LLMs in cybersecurity knowledge using retrieval-augmented generation. However, it lacks practical IOC classification techniques and remains limited to Q\&A format testing rather than real-world threat detection scenarios.

CyberLLMInstruct \cite{cyberLLMInstruct} developed 54,928 instruction-response pairs for identifying cybersecurity risks including malware, phishing, and zero-day vulnerabilities. Research shows fine-tuning on cybersecurity data creates significant performance-safety tradeoffs, with models experiencing substantial safety degradation (e.g., Llama 3.1 8B security scores dropping from 0.95 to 0.15). Neither dataset provides comprehensive techniques for automated IOC classification from unstructured threat intelligence sources.

\subsection{Specialized Threat Detection Algorithms}

Zhang et al. \cite{phishing} developed phishing detection for small and medium enterprises, achieving notable performance but remaining constrained to phishing threats with limited generalizability beyond SME environments. PMANet \cite{pmanet} focuses exclusively on malicious URL detection through post-trained language models, exemplifying the problem of providing solutions for individual threat types rather than comprehensive IOC detection.



\subsection{Traditional Machine Learning Approaches}

Conventional supervised and unsupervised learning approaches for malware detection \cite{superUnsuper}, \cite{10830310}, \cite{10775101}, \cite{10414983}, \cite{10048243} represent the predominant paradigm in current systems. These approaches suffer from fundamental limitations as reactive systems that detect threats only after observation and cataloging. Key limitations include inability to detect zero-day attacks due to reliance on known signatures and poor adaptability to evolving malware families using polymorphic techniques.

\subsection{Research Gaps and Our Contribution}

This work addresses the following gaps in existing approaches:

\begin{enumerate}
    \item \textbf{Lack of Proactive Detection}: Most approaches operate reactively i.e. detecting threats after compromise.
    \item \textbf{Limited Multi-Source Integration}: Systems typically process single source types, lacking unified approaches for diverse threat intelligence
    \item \textbf{Inadequate Context-Aware Classification}: Methods struggle with contextual understanding needed to distinguish malicious from benign indicators
    \item \textbf{Evaluation Limitations}: Limited systematic evaluation across multiple LLM architectures and real-world sources
    \item \textbf{Scalability Challenges}: Manual processes that do not scale to modern threat intelligence volume
\end{enumerate}
\section{Data Collection}
\begin{figure}[!tbp] 
    \centering 

    \begin{subfigure}[t]{0.48\columnwidth} 
        \centering
        \includegraphics[width=\linewidth, height=1in]{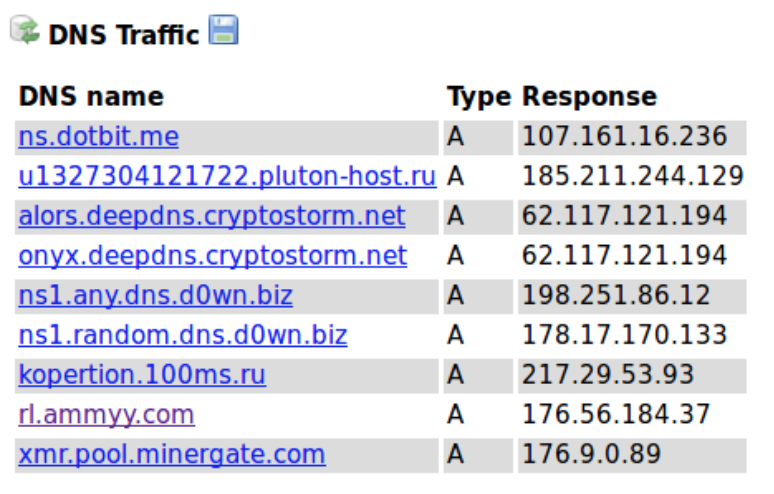}
        \caption{IOC data in image}
        \label{fig:ioc_image_side} 
    \end{subfigure}
    \hfill 
    \begin{subfigure}[t]{0.48\columnwidth} 
        \centering
        \includegraphics[width=\linewidth]{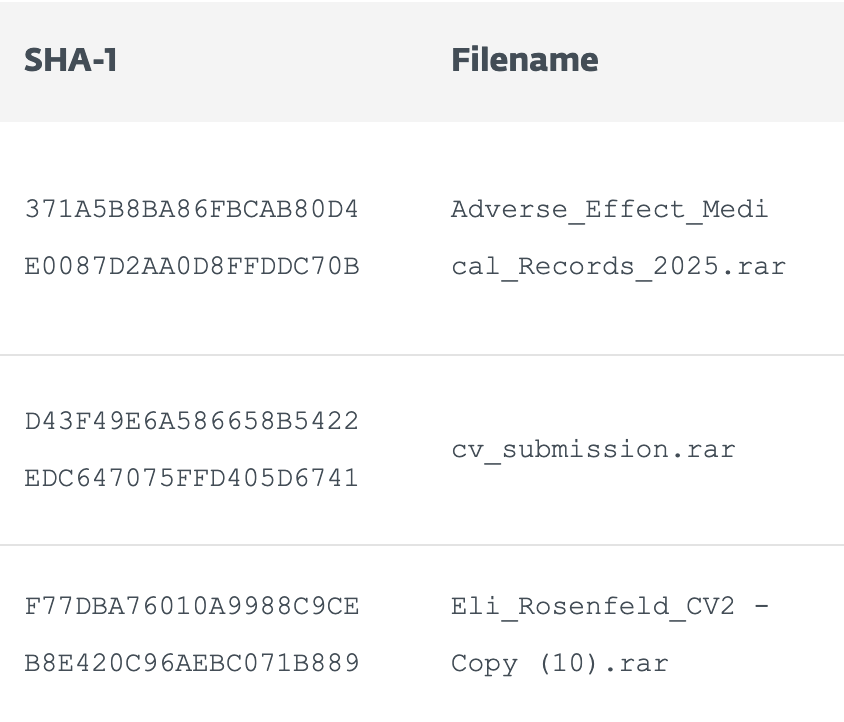}
        \caption{IOC data in hash format}
        \label{fig:ioc_hash_side} 
    \end{subfigure}

    \caption{Examples of different IOC data representations (side-by-side) \cite{abusech}}
    \label{fig:side_by_side_ioc_examples} 
\end{figure}

We started with a list of threat report RSS feeds published by popular cybersecurity companies like Sophos, SentinelOne, Crowdstrike, Mandiant, Abuse.ch etc. Our automated system consists of a web crawler that stores the raw page data from each of the posts published to the threat report feeds. An example of one of these websites is in Fig.\ref{fig:side_by_side_ioc_examples}. This shows that even within a single website -- in this case Abuse's \cite{abusech} RSS feed -- there are multiple ways and types of indicator data hosted on the webpages. As there is no standardized way of making the IOC data available, every report publisher decides to use a different format, making it extremely hard to have an easily parsable database. Moreover, attempts to standardize threat intelligence sharing like STIX 2.0 \cite{stix2} haven't been adopted across the industry, especially for time-sensitive, cutting-edge intelligence.

Our system then parses the raw page data from each of these web pages using regular expressions to extract Indicators of Compromise (IOCs). The regular expressions attempt to find patterns that look like hashes (SHA1, MD5, SHA256), IPv4 addresses, IPv6 addresses, domains and URLs. In our analysis, we focus on the text information from a web oage only. The raw data collected is stored in a search index for quick retrieval. We also store the metadata from each of the raw pages for easy lookup.

The parser also uses the MITRE ATT\&CK data \cite{mitreattack} Python library to extract patterns that look like well known Tactics, Techniques and Groups. The data is in STIX 2.0 \cite{stix2} format and constitute a knowledge base of observed adversary tactics and techniques. It is structured as a collection of objects that represent various concepts related to cyber adversary behavior.
\section{Threat Intelligence Sources \& Analysis}
\subsection{Threat Intelligence Sharing Formats}
The cybersecurity community has standardized threat intelligence sharing through STIX (Structured Threat Information Expression) and TAXII (Trusted Automated eXchange of Intelligence Information) \cite{stix2} frameworks . These OASIS standards create a common language for exchanging cyber threat intelligence in structured, machine-readable formats.

However, industry-wide STIX/TAXII adoption remains fragmented due to specification complexity and implementation challenges. Analysis of approximately 6 million STIX objects over nine years reveals that security providers generate only 2,063 unique daily objects, inadequate for increasing cyber threats \cite{Borgolte2024}. Additionally, 37.89\% of STIX objects show substantial redundancy even from single providers.

Consequently, free-structured threat reports dominate the CTI landscape, with 69\% of teams relying on news and media sources according to the SANS 2023 CTI Survey \cite{Silobreaker2023}.

\subsection{The role of AI and Automation}

Artificial intelligence and automation are transforming threat intelligence processing. The SANS 2024 Detection and Response Survey indicates 87\% of organizations use automated threat detection tools, while 67\% plan to expand AI/ML capabilities \cite{Prelude2024}. This automation trend addresses challenges from high-volume unstructured data and resource-intensive manual processing.

\subsection{Threat Intelligence News Sources}

Effective AI threat detection requires high-quality intelligence sources. We evaluated dozens of sources against three criteria to select 15 providers:

\begin{enumerate}
  \item Reputation and expertise from established cybersecurity firms or government agencies offering higher-fidelity IOCs
  \item Technical depth providing contextual analysis linking IOCs with tactics, techniques, and procedures (TTPs)
  \item Diversity of indicator types including hashes, IP addresses, domains, threat actors, and vulnerabilities for richer contextual understanding
\end{enumerate}

\subsection{Web-crawler and Parser}

The algorithm for crawling and parsing threat report web pages is outlined below:


\begin{enumerate}
  \item Retrieve threat report URLs from RSS/Atom feeds
  \item Extract webpage content and textual bodies
  \item Identify IOCs (IP addresses, domains, hashes, CVEs, YARA rules) using pattern-matching and defanging techniques
  \item Enrich reports by mapping terms to MITRE ATT\&CK framework entities
  \item Store structured IOCs, ATT\&CK entities, and metadata in databases
  \item Index textual content for efficient full-text searching of collected threat intelligence
\end{enumerate}
\section{Performance of LLMs in IOC Classification}
For our studies, we collected 479 web pages containing 711 IPv4 addresses, 502 IPv6 addresses, and 1445 domains that need to be analyzed. We focused on IP addresses and domains since SHA1, SHA256 and MD5 hashes show very low or zero false positives due to their relative well-structured nature. We created a smaller subset to test, consisting of 303 web pages that contained 116 IPv4 addresses and 187 domains that we analyzed. Of these 303 indicators, 251 were malicious. We categorized them as malicious based on both the webpage context as well as searching against malware databases like VirusTotal to ensure they are actually malicious.

In this evaluation, we asked the model to classify an Indicator of Compromise into its correct category (IPv4 address, domain name etc). We used the pre-trained LLM  models without specific fine-tuning for this task. We created the following prompt that was sent along with the question to identify if the Indicator is an IOC or not.

"\textit{You are the best cybersecurity expert. You can detect all Indicators of Compromise. Your Task is to only answer the following question in true or false for the question provided. Do not add any more sentences. You have to use the following context. \\ Context: \{context\}\\ Question: Is the  \textless IOC Type\textgreater\ \textless IOC\textgreater\ an Indicator of Compromise?'}"

The context window consists of the whole webpage in text format. The system removed all the HTML tags and sent the scraped webpage to the model.

\begin{table}[!t]
\caption{Confusion Matrix for Gemini Models}
\label{tab:confusion_matrix_flash_pro}
\centering
\begin{tabular}{@{}llcccc@{}}
\toprule
& & \multicolumn{4}{c}{\textbf{Predicted Class}} \\
\cmidrule(l){3-6} 
& & \multicolumn{2}{c}{\textbf{Gemini 2.0 Flash-Lite}} & \multicolumn{2}{c}{\textbf{Gemini 1.5 Pro}} \\
\cmidrule(l){3-4} \cmidrule(l){5-6}
& & Malicious & Benign & Malicious & Benign \\
\midrule
\multirow{2}{*}{\rotatebox[origin=c]{90}{\textbf{Actual}}} & Malicious & 246 & 5 & 251 & 0 \\
& Benign    & 16  & 36 & 11  & 41 \\
\\
\bottomrule
\end{tabular}
\end{table}



\begin{table}[!t]
\caption{Confusion Matrices for Qwen3 Models}
\label{tab:confusion_matrix_qwen}
\centering
\begin{tabular}{@{}llcccc@{}}
\toprule
& & \multicolumn{4}{c}{\textbf{Predicted Class}} \\
\cmidrule(l){3-6} 
& & \multicolumn{2}{c}{\textbf{Qwen3 30B}} & \multicolumn{2}{c}{\textbf{Qwen3 32B}} \\
\cmidrule(l){3-4} \cmidrule(l){5-6}
& & Malicious & Benign & Malicious & Benign \\
\midrule
\multirow{2}{*}{\rotatebox[origin=c]{90}{\textbf{Actual}}} & Malicious & 170 & 81 & 239 & 12 \\
& Benign    & 31  & 21 & 51  & 1  \\
\\
\bottomrule
\end{tabular}
\end{table}



\begin{table}[!t]
\caption{Confusion Matrices for Llama Models}
\label{tab:confusion_matrix_llama}
\centering
\begin{tabular}{@{}llcccc@{}}
\toprule
& & \multicolumn{4}{c}{\textbf{Predicted Class}} \\
\cmidrule(l){3-6} 
& & \multicolumn{2}{c}{\textbf{Llama 8B}} & \multicolumn{2}{c}{\textbf{Llama 70B}} \\
\cmidrule(l){3-4} \cmidrule(l){5-6}
& & Malicious & Benign & Malicious & Benign \\
\midrule
\multirow{2}{*}{\rotatebox[origin=c]{90}{\textbf{Actual}}} & Malicious & 214 & 37 & 251 & 0  \\
& Benign    & 23  & 29 & 18  & 34 \\
\\
\bottomrule
\end{tabular}
\end{table}



The Gemini models (2.0 flash-lite and 1.5 Pro) demonstrated high Recall for malicious indicators. 
This suggests that they are effective at capturing a large proportion of actual threats. However, Gemini 1.5 Pro performance in correctly identifying non-malicious indicators (Specificity) was notably higher. 
This is also evident in the higher Precision rate(0.958) for malicious IOCs for Gemini models, 
which, while high, is impacted by these false positives.

Qwen3 32b, on the other hand, showed the worst performance with a relatively lower number of true positives (239) and low false positives (1) when identifying malicious indicators, and high false negatives (51) when classifying non-malicious data. This resulted in a low F1-score(the harmonic mean of Precision and Recall) for this model. 

The Llama 70B model, while identifying all malicious indicators (TP=251), had a higher false negatives (FN=18) compared to some Gemini models, resulting in a lower Recall for the malicious class. 

These metrics underscore the trade-offs inherent in classification: a model excelling at catching all malicious instances (high Recall) might do so at the cost of incorrectly flagging more benign instances (lower Precision for malicious, lower Specificity for non-malicious ) and vice-versa. For proactive threat detection, a high Recall for malicious IOCs is paramount, but a reasonable precision is also needed to ensure the system is usable by security analysts.

\subsubsection{When Models Misidentify Benign Data}
In this section we discuss the model's evaluation on the non-malicious data. The Gemini models were good in identifying a good percentage of malicious indicators but they performed poorly in identifying the non-malicious indicators. As an example,
one of the indicator types that Gemini classified as malicious is \textit{http://\textless C\&C\_IP\textgreater :\textless C\&C\_port\textgreater /anydesk.exe}. A human can quickly identify this as an example of a potentially malicious domain, not an actual malicious indicator itself.

In our experiments, we observed that models do not perform well on domain type indicators. We hypothesize that, in addition to the context, the models consider the domain names themselves when evaluating their overall malicious intent. As indicated by certain domain-type features identified as malicious for ex.: \textit{TrueSightKiller} in \url{github.com/MaorSabag/TrueSightKiller}, \textit{abuse} in \url{https://abuse.ch/downloads/blog/adwind\_domains\_20170828.txt}, \textit{Darkside} in \url{github.com/ph4nt0mbyt3/Darkside} etc.



\section{Discussion}
\begin{figure}
    \centering
    \includegraphics[width=0.9\linewidth]{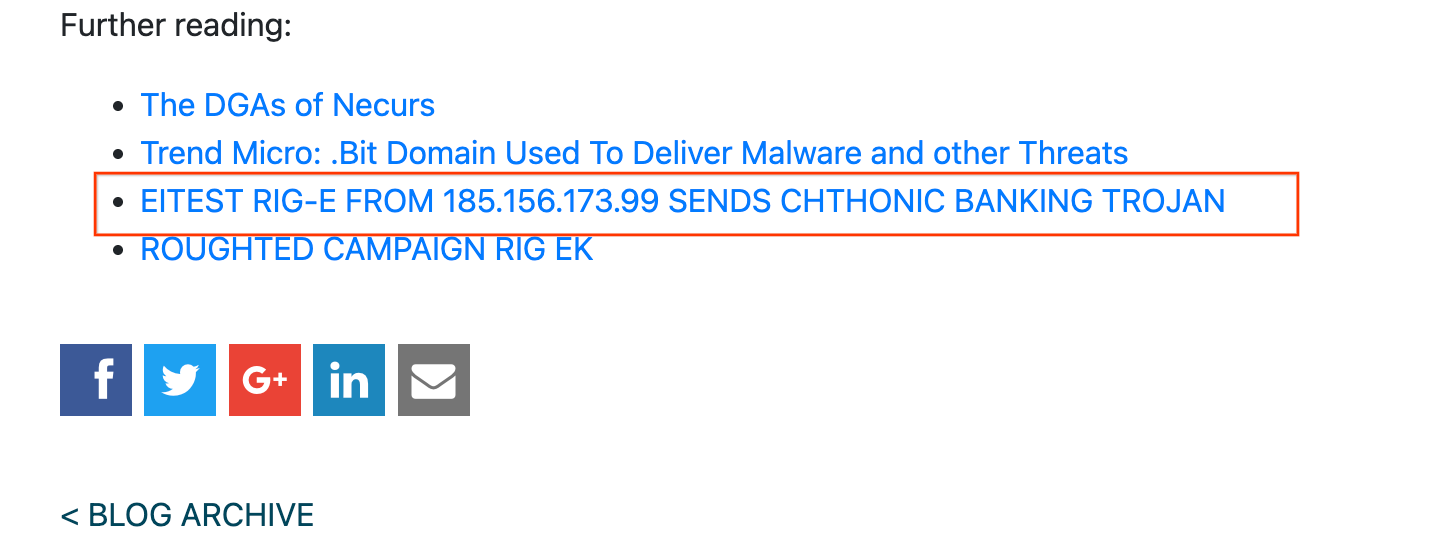}
    \caption{IP address that looks malicious from context \cite{abusech2011}.}
    \label{fig:ioc-identify}
\end{figure}

The experiments instructed models to rely solely on parsed webpage context to limit hallucination and maintain consistency. This may have caused some unintentional consequences for the overall results. For example, the IP address \textit{185.156.173.99} shown in Fig.\ref{fig:ioc-identify} is tagged malicious by the models. This is found in the \textit{Further Reading} section of the website, and is not directly indicative of compromise.

Domain registrars such as GoDaddy and Squarespace employ sophisticated security measures, but less reputable registrars often appear as malicious in contexts since they host malware sites; this leads to false positives as the registrars themselves are not malicious. Distinguishing between malicious domains and registrar context requires fine-tuning with specific data.

Overall, models demonstrated difficulties accurately classifying domains and non-malicious indicators, highlighting challenges in contextual understanding and the need for improved model refinement and data specificity for effective threat detection.

\section{Future Work}

While this paper demonstrates LLMs' ability to provide cutting-edge performance on malicious indicator identification through contextual understanding, several expansion opportunities exist for enhancing the approach. Future research can extend beyond HTML threat reports to parse diverse unstructured and semi-structured document formats including PDF, Microsoft Word, and RTF files, while incorporating image analysis capabilities to extract malicious indicators from screenshots of popular tools like Wireshark \cite{wireshark} that are commonly included in threat reports but absent from textual content. Additionally, the methodology can be expanded to identify more complex indicator types beyond IPv4 addresses and domains, such as tactics, techniques, attack groups, nation states, file names, program versions, registry keys, mutexes, and other critical indicators of compromise that pose greater parsing challenges due to their semantic ambiguity. For instance, common English words like Panda Stealer require contextual interpretation to be identified as malware families. The proposed approach is expected to generalize well to these difficult-to-parse indicator types, offering comprehensive threat intelligence automation capabilities for enterprise security systems.

\bibliographystyle{unsrt}
\bibliography{main}

\end{document}